\documentclass[twocolumn,twoside,preprintnumbers,amsmath,showpacs,amssymb]{revtex4}
\usepackage{amsfonts,amssymb,amsmath}
\usepackage{graphicx}% Include figure files
\usepackage{dcolumn}% Align table columns on decimal point
\usepackage{fancyhdr}

\newcommand{\be}{\begin{equation}}
\newcommand{\bea}{\begin{eqnarray}}
\newcommand{\ee}{\end{equation}}
\newcommand{\eea}{\end{eqnarray}}
\newcommand{\bpi}{\begin{picture}}
\newcommand{\bce}{\begin{center}}
\newcommand{\epi}{\end{picture}}
\newcommand{\ece}{\end{center}}
\newcommand{\D}{\displaystyle}
\def\chic#1{{\scriptscriptstyle #1}}

\def\g{\widetilde{{\rm I}\hspace{-0.07cm}\Gamma}}
\def\gt{\widetilde{\Gamma}^{\rm L}_{\nu\alpha\beta}}
\def\gnp{{\overline g}^2_{{\chic {\rm NP}}}}
\newcommand{\Valencia}{Departamento de F\'\i sica Te\'orica
and IFIC, Centro Mixto, Universidad de Valencia--CSIC \\
E-46100, Burjassot, Valencia, Spain}

\begin{document}

\title{{\Large Analyzing dynamical gluon mass generation }}

\author{Arlene C. Aguilar}
\email{aguilar@ift.unesp.br}
\affiliation{Instituto de F\'{\i}sica Te\'orica,
Universidade Estadual Paulista,
Rua Pamplona 145,
01405-900, S\~ao Paulo, SP,
Brazil }

\author{Joannis Papavassiliou}
\email{Joannis.Papavassiliou@uv.es}
\affiliation{\Valencia}

\received{}

\begin{abstract}

We  study  the  necessary  conditions for  obtaining  infrared  finite
solutions from the Schwinger-Dyson  equation governing the dynamics of
the gluon propagator. The equation in question is set up in the Feynman
gauge  of the  background field  method,  thus capturing  a number  of
desirable  features.  Most  notably, and  in contradistinction  to the
standard   formulation,   the    gluon   self-energy   is   transverse
order-by-order  in  the dressed  loop  expansion,  and separately  for
gluonic  and  ghost   contributions.   Various  subtle  field-theoretic
issues, such as renormalization group invariance and regularization of
quadratic  divergences,  are  briefly  addressed.   The  infrared  and
ultraviolet  properties  of the  obtained  solutions  are examined  in
detail,  and  the  allowed  range  for the  effective  gluon  mass  is
presented.

% 1st  Other nonperturbative calculations
% 2nd  General properties of QCD (dynamics, confinement, etc)

\end{abstract}

\maketitle

\thispagestyle{fancy}
\setcounter{page}{0}

The  most widely  used  approach  for studying  in  the continuum  QCD
effects  that lie  beyond the  realm  of perturbation  theory are  the
Schwinger-Dyson  (SD)  equations.   This  infinite system  of  coupled
non-linear integral equations for  all Green's functions of the theory
is inherently  non-perturbative and can accommodate  phenomena such as
chiral symmetry  breaking and dynamical mass  generation.  In practice
one  is  of  course  severely   limited  in  their  use,  and  various
approximations have been implemented throughout the years.  Devising a
self-consistent  truncation  scheme for  the  SD  series  is far  from
trivial. The main problem in this context is that the SD equations are
built  out of unphysical  Green's functions;  thus, the  extraction of
reliable physical information  depends crucially on delicate all-order
cancellations, which  may be inadvertently distorted in  the process of
the truncation.   In order  to partially compensate  for this  type of
shortcomings, one  usually attempts to supplement  as much independent
information as possible,  by ``solving" the complicated Slavnov-Taylor
identities  (STI),   or  by   combining  with  results   from  lattice
simulations.

The   truncation   scheme   based   on  the   pinch   technique   (PT)
~\cite{Cornwall:1982zr,Cornwall:1989gv} implements     a     drastic
modification already  at the  level of the  building blocks of  the SD
series, namely the off-shell  Green's functions themselves.  The PT is
a well-defined  algorithm that exploits  systematically the symmetries
built into physical observables, such as $S$-matrix elements, in order
to  construct  new,  effective  Green's functions  endowed  with  very
special  properties.  Most  importantly, they  are independent  of the
gauge-fixing parameter, and  satisfy naive (ghost-free, QED-like) Ward
identities (WI) instead  of the usual STI. The  upshot of this program
is to first  trade the conventional SD series  for another, written in
terms of  these new Green's  functions, and subsequently  truncate it,
keeping only a few  terms in a ``dressed-loop'' expansion, maintaining
at the same time exact gauge-invariance.

Of central importance in this context is the connection between the PT
and  the Background  Field  Method  (BFM).  The  latter  is a  special
gauge-fixing procedure that preserves the symmetry of the action under
ordinary  gauge   transformations  with  respect   to  the  background
(classical) gauge field $\widehat{A}^a_{\mu}$, while the quantum gauge
fields  $A^a_{\mu}$  appearing in  the  loops transform  homogeneously
under  the  gauge  group \cite{Abbott:1980hw}. As  a  result,  the 
background  $n$-point functions satisfy  QED-like all-order WIs. 
The connection
between PT and  BFM, known to persist to all  orders, affirms that the
(gauge-independent) PT effective $n$-point functions coincide with the
(gauge-dependent) BFM $n$-point functions provided that the latter are
computed in the Feynman gauge \cite{Binosi:2002ft}.

In this talk  we consider the all-order diagrammatic  structure of the
effective   gluon  self-energy,   $\widehat\Pi_{\mu\nu}(q)$,  obtained
within  the PT-BFM  framework. We  explain that,  as a  consequence of
all-order  WI  satisfied  by   the  full  vertices  appearing  in  the
corresponding  diagrams, the
transversality  of  $\widehat\Pi_{\mu\nu}(q)$ is  realized  in a  very
special way:  the contributions  of gluonic and  ghost loops  are {\it
separately transverse}. In particular, we study a truncated version of
this  new series,  keeping  only  the terms  of  the gluonic  one-loop
dressed expansion,  while still maintain exact  gauge invariance.  Our
attention  will  focus  on   a  detailed  scrutiny  of  the  necessary
conditions  for  obtaining  infrared  finite  solutions  from  the  SD
equation.

Let us first define  
some basic quantities. 
First of all, it should be clear from the beginning that
there are two different gluon propagators appearing in this problem,
$\widehat{\Delta}_{\mu\nu}(q)$, denoting the background gluon propagator, and ${\Delta}_{\mu\nu}(q)$, denoting 
the quantum gluon propagator appearing inside the loops.
In the  Feynman gauge, $\widehat{\Delta}_{\mu\nu}(q)$ is given by 
\begin{equation}
\widehat{\Delta}_{\mu\nu}(q)= {-\D i}\left[{\rm P}_{\mu\nu}(q)\widehat{\Delta}(q^2) + 
\frac{q_{\mu}q_{\nu}}{q^4}\right],
\label{prop_cov}
\end{equation}
where the transversal projector, 
\be
{\rm P}_{\mu\nu}(q)= \ g_{\mu\nu} - \frac{\D q_\mu
q_\nu}{\D q^2}.
\ee
The scalar function $\widehat{\Delta}(q^2)$ is related to the 
all-order gluon self-energy $\widehat{\Pi}_{\mu\nu}(q)$ by
\be
\widehat{\Pi}_{\mu\nu}(q)={\rm P}_{\mu\nu}(q)\,\widehat{\Pi}(q^2)\,; \quad \widehat{\Delta}(q^2) = \frac{1}{q^2 + i\widehat{\Pi}(q^2)}\,.
\ee
Exactly analogous definitions relate ${\Delta}_{\mu\nu}(q)$ 
with $\Pi_{\mu\nu}(q)$.

\begin{figure}[ht]
\includegraphics[scale=0.65]{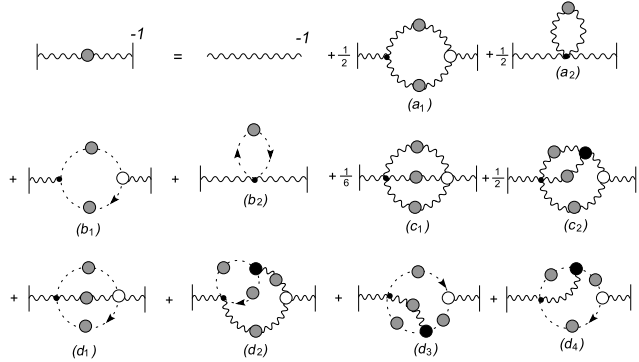}
\caption{The SD equation for the gluon propagator. Wavy lines with grey blobs represent
full-quantum gluon propagators, while the dashed lines with 
grey blobs denote full-ghost propagators. All external wavy lines (ending with a
vertical line) are background gluons.  The black dots are
the  tree-level  vertices  in  the  BFM,  while black blob represents the
 full conventional  vertices. The white blobs 
denote three or four-gluon vertices with one external background leg.}
\label{f1}
\end{figure}

The            diagrammatic          representation           of
$\widehat{\Delta}_{\mu\nu}^{\,-1}(q)$        is        shown        in
Fig.(\ref{f1}). Notice  that diagrams  $(b_2)$, $(d_1)$, $(d_2)$,
$(d_3)$ and  $(d_4)$, are characteristic 
to the BFM; within the PT they are generated {\it dynamically}, from the  
STIs triggered by the pinching momenta.  

%\be
% \tilde b = \frac{10 C_A}{48\pi^2}
%\label{bfactor}
%\ee

%%%%%%%%%%%%%%%%%%%%%%%% FIGURE 3 %%%%%%%%%%%%%%%%%%%%%%%%%%%%%%%%%%%%%
%           One-loop dressed ghost contribution - Group B
%%%%%%%%%%%%%%%%%%%%%%%%%%%%%%%%%%%%%%%%%%%%%%%%%%%%%%%%%%%%%%%%%%%%
%%%%%%%%%%%%%%%%%%%%%%%%%%%%%%%%%%%%%%%%%%%%%%%%%%%%%%%%%%%%%%%%%%%%%%%

As is  widely known, in the conventional formalism  
the inclusion of ghosts is instrumental for  the   transversality   of
$\Pi^{ab}_{\mu\nu}(q)$, already at the level of the one-loop calculation.
On  the other  hand, in the  PT-BFM formalism,
due  to new Feynman  rules for  the vertices,  the one-loop  gluon and
ghost contribution are individually transverse \cite{Abbott:1980hw}.

As has been shown in \cite{Aguilar:2006gr},
this crucial feature  persists  at the 
non-perturbative level, as a consequence of the simple WIs satisfied by
the full vertices appearing in
the  diagrams  of Fig.(\ref{f1}).
Specifically, the  gluonic  and  ghost  sector  are
separately  transverse, within each individual order in the dressed-loop expansion

We will show this property for the one-loop dressed terms.
We start by writing down the fundamental all-order WI for
the full three-gluon vertex with one background gluon, ${\g}_{\mu\alpha\beta}^{abc}$,
and for the full background gluon-ghost vertex ${\g}_{\mu}^{acb}$,
\bea
q_1^{\mu}{\g}_{\mu\alpha\beta}^{abc}(q_1,q_2,q_3) &=&
gf^{abc}
\left[\Delta^{-1}_{\alpha\beta}(q_2)
- \Delta^{-1}_{\alpha\beta}(q_3)\right] \,,
\nonumber\\
q_1^{\mu}{\g}_{\mu}^{acb}(q_2,q_1,q_3) &=&  gf^{abc}
\left[D^{-1}(q_2)- D^{-1}(q_3)\right] \,,
\label{3gl} 
\eea
where on the RHS we have differences of inverse of the quantum gluon, $\Delta_{\mu\nu}(q)$, and ghost, $D(q)$, propagators.  

The closed expressions 
corresponding to the gluonic sector, at one-loop dressed expansion, (see Fig.(\ref{f1})) are given by
\begin{eqnarray}
\widehat{\Pi}^{ab}_{\mu\nu}(q)
\big|_{{\bf a_1}} &=& 
\frac{1}{2} \, \int\!\! [dk]\,
\widetilde{\Gamma}_{\mu\alpha\beta}^{aex}
\Delta^{\alpha\rho}_{ee'}(k)
{\g}_{\nu\rho\sigma}^{be'x'}
\Delta^{\beta\sigma}_{xx'}(k+q)  \,,
\nonumber\\
\widehat{\Pi}^{ab}_{\mu\nu}(q)
\big|_{{\bf a_2}} &=&
\frac{1}{2} \,\int\!\! [dk]\,
\widetilde{\Gamma}_{\mu\nu\alpha\beta}^{abex}
\Delta^{\alpha\beta}_{ex} (k) \, ,
\label{groupa}
\end{eqnarray}
with $\widetilde{\Gamma}_{\mu\alpha\beta}^{aex}$ and  $\widetilde{\Gamma}_{\mu\nu\alpha\beta}^{abex}$ being the three and four bare gluon vertices  
in the Feynman gauge of the BFM \cite{Abbott:1980hw}.

For the ghost sector, we have
\begin{eqnarray}
\widehat{\Pi}^{ab}_{\mu\nu} (q)
\big|_{{\bf b_1}} &=& 
- \, \int\!\! [dk]\,
\widetilde{\Gamma}_{\mu}^{aex}
D_{ee'}(k) 
{\g}_{\nu}^{be'x'}
D_{xx'}(k+q)  \,,
\nonumber\\
\widehat{\Pi}^{ab}_{\mu\nu}(q)
\big|_{{\bf b_2}} &=&
- \int\!\! [dk]\,
\, \widetilde{\Gamma}_{\mu\nu}^{abex}
D_{ex}(k)\,.
\label{groupb}
\end{eqnarray}
where $\widetilde{\Gamma}_{\mu}^{aex}$ and $\widetilde{\Gamma}_{\mu\nu}^{abex}$
represent the tree-level ghost-gluon vertices with one (two) background gluon(s) 
respectively \cite{Abbott:1980hw}; the measure $ [dk] =  d^d k/(2\pi)^d$ \, 
with $d=4-\epsilon$ the dimension of space-time.
Observe that in our notation all the three and four-point functions with a tilde are vertices with at least one external (background) gluon leg.

With the above WI we can prove that the groups (a) and (b) are independently transverse. We start with group (a)
\bea
q^{\nu} \widehat{\Pi}^{ab}_{\mu\nu}(q)
\big|_{{\bf a_1}} &=&
 C_A \, g^2 \delta^{ab} \,q_{\mu}  \, \int\!\! [dk]\,
\Delta^{\rho}_{\rho}(k) \,,
\nonumber\\
q^{\nu}\widehat{\Pi}^{ab}_{\mu\nu}(q)
\big|_{{\bf a_2}} &=&
- C_A \, g^2  \delta^{ab} \,q_{\mu}  \, \int\!\! [dk]\,
\Delta^{\rho}_{\rho}(k) \,,
\label{Transgroupa}
\eea
and thus 
\be
q^{\nu}\left( 
\widehat{\Pi}^{ab}_{\mu\nu}(q)\big|_{{\bf a_1}}+
\widehat{\Pi}^{ab}_{\mu\nu}(q)\big|_{{\bf a_2}}\right) = 0  \,.
\ee

Similarly, the one-loop-dressed ghost contributions give upon
contraction
\bea
q^{\nu} \widehat{\Pi}^{ab}_{\mu\nu}(q)
\big|_{{\bf b_1}} &=&
2 \,C_A \, g^2 \delta^{ab} \,q_{\nu}  \, \int\!\! [dk]\,D(k)  \,,
\nonumber\\
q^{\nu} \widehat{\Pi}^{ab}_{\mu\nu}(q)
\big|_{{\bf b_2}} &=& -2 \,C_A \, g^2 \delta^{ab} 
\,q_{\nu}  \, \int\!\! [dk]\,D(k) \,,
\label{Transgroupb}
\eea
and so
\be
q^{\nu}\left( 
\widehat{\Pi}^{ab}_{\mu\nu}(q)\big|_{{\bf b_1}}+
\widehat{\Pi}^{ab}_{\mu\nu}(q)\big|_{{\bf b_2}}\right) = 0  \,.
\ee

The proof of the individual transversality 
of the groups (c) and (d), constituting the two-loop dressed expansion of  $\widehat{\Pi}_{\mu\nu}(q)$,
is slightly more cumbersome but essentially straightforward~\cite{Aguilar:2006gr}.

 The importance of  this transversality property in the  context of SD
equation  is that  it  allows for  a  meaningful first  approximation:
instead of the  system of coupled equations involving  gluon and ghost
propagators,  one  may consider  only  the  subset containing  gluons,
without  compromising the  crucial property  of  transversality.  More
generally,  one  can envisage  a  systematic  dressed loop  expansion,
maintaining transversality  manifest at every  level of approximation.
This is  not to say,  of course, that  we have some a-priori guarantee
that the  subset of diagrams considered here  is numerically dominant.
Actually, as has been argued in a series of SD studies, in the context
of the conventional Landau gauge it is the ghost sector that furnishes
in fact  the leading contribution~\cite{vonSmekal:1997is}  Clearly, it
is plausible  that this characteristic feature may  persist within the
PT-BFM scheme as  well, and we will explore this  crucial issue in the
near future

Thus,  in  this formalism,  the  first  non-trivial approximation  for
$\widehat{\Delta}^{-1}(q^2)$  that  preserves  its  transversality  is
given by the  gluonic terms of the one-loop expansion
(diagrams ($a_1$) and ($a_2$) in Fig.(\ref{f1})),
written in closed form in Eq.~(\ref{groupa}).

 However, the equation given in (\ref{groupa}) is not a genuine SD equation, in the sense
that it does not involve the unknown quantity $\widehat{\Delta}$ on both sides;
instead, in the integrals of the RHS appears $\Delta$.
Replacing $\Delta$ by $\widehat{\Delta}$ is a highly non-trivial proposition,
whose self-consistency is still an open issue.
Its implementation may be systematized by resorting to a set of crucial identities relating
$\Delta$ and $\widehat{\Delta}$ by means of a set of auxiliary Green's functions
involving anti-fields and background sources. At this point we will 
assume that to a first approximation one may neglect
the effects of the aforementioned auxiliary Green's functions, 
and carry out the substitution $\Delta \to \widehat{\Delta}$ on the RHS of (\ref{groupa}).
Hence, the SD equation we will solve is 
written as
\bea
\widehat{\Delta}^{-1}(q^2)
=  q^2  + i \left[\widehat{\Pi}^{(\bf{a_1})}(q^2) + 
 \widehat{\Pi}^{(\bf{a_2})} \right] \,,
 \label{sd1}
 \eea
where 
 \bea
\widehat{\Pi}_{\mu\nu}^{(\bf{a_1})}(q) &=& 
\frac{1}{2}\, C_A \, g^2 \, \int\!\! [dk]
\widetilde{\Gamma}_{\mu\alpha\beta}
\widehat{\Delta}^{\alpha\alpha'}(k) 
{\g}_{\nu\alpha'\beta'} 
\widehat{\Delta}^{\beta\beta'}(k+q)  \,, \nonumber\\
\widehat{\Pi}_{\mu\nu}^{(\bf{a_2})}  &=&  - C_A \, g^2 \,g_{\mu\nu} 
\int\!\! [dk] \widehat{\Delta}(k) \,.
\label{polar}
\eea

With the  above equation at hand, we proceed to establish
the conditions necessary for 
obtaining infrared  finite solutions for $\widehat{\Delta}^{-1}(q^2)$, i.e.
solutions  for   which  $\widehat{\Delta}^{-1}(0)\neq  0$.
There are two such conditions: one must (i) allow for non-vanishing seagull-like contribution, and
(ii) introduce massless poles into the Ansatz for the full three-gluon
vertex.

The necessity of the first condition can be appreciated by 
observing that, on dimensional grounds, 
the value of  $\widehat{\Delta}^{-1}(0)$  
can only be proportional to two types of seagull-like contributions, 
\be
{\cal T}_{0} = \int\!\,[dk]\,\widehat\Delta(k), \qquad
 {\cal T}_{1} = \int\!\,[dk]\,k^2 \, \widehat\Delta^{2}(k) \,,
\label{seagullgen}
\ee 
since, inside the diagrams  of Eq.(\ref{polar}), there can be 
at most  two full gluon  self-energies, $\widehat{\Delta}$.  However,
it is  well known  that, due to  the dimensional  regularization rules,
such contributions  vanish perturbatively, ensuring  the masslessness
of the gluon order by order in perturbation theory.  In order 
for finite  solutions to emerge,  one   must  assume  that
seagull-like contributions, such as those of Eq.(\ref{seagullgen}), do
not vanish  non-perturbatively.  Naturally, this last  step will force
us to deal  with the quadratic divergences, present  in both integrals
of  Eq.(\ref{seagullgen}),  and  therefore a  suitable  regularization
scheme must be subsequently employed.

Allowing the non-vanishing of seagull-like terms is not the whole story 
however;  one must determine in addition the mechanism that will 
produce their appearance. One thing is certain:  
the seagull contributions determining $\widehat{\Delta}^{-1}(0)$ 
do {\it not} originate from diagram $(a_2)$ in Fig.(\ref{f1}).
Instead, the required seagull contributions will 
stem from diagram $(a_1)$, {\it after}  the inclusion of massless
poles into the Ansatz for the full three-gluon vertex $\g$.
Diagram $(a_2)$ plays of course a crucial role in 
enforcing the transversality of  $\widehat{\Pi}_{\mu\nu}$ non-perturbatively,
but in the absence of massless poles in the vertex one would still get
$\widehat{\Delta}^{-1}(0)=0$.

There is a relatively simple argument that amply demonstrates the subtle
interplay between both  requirements.  
Specifically, $i\widehat{\Pi}_{\mu\nu}^{(\bf{a_1})}(q)$  can be
  written in the general form
\be
i\widehat{\Pi}_{\mu\nu}^{(\bf{a_1})}(q)= 
q^2 A(q^2) g_{\mu\nu} + B(q^2) q_{\mu} q_{\nu}\,,
\label{genform}
\ee
where $A(q^2)$ and $B(q^2)$ are arbitrary dimensionless functions, 
whose precise expressions depend on the details of the 
${\g}_{\nu\alpha'\beta'}$ employed. 
The transversality of $i \left[ \widehat{\Pi}_{\mu\nu}^{(\bf{a_1})}(q) + 
 \widehat{\Pi}_{\mu\nu}^{(\bf{a_2})} \right]$
implies immediately the 
condition 
\be
q^2 \left[ A(q^2)+B (q^2)\right]  =  i C_A \, g^2 \,  {\cal T}_{0} \, ,
\ee
and thus the sum of the two graphs reads  
\be
i \left[ \widehat{\Pi}_{\mu\nu}^{(\bf{a_1})}(q) + 
 \widehat{\Pi}_{\mu\nu}^{(\bf{a_2})} \right]
= - q^2 B(q^2) P_{\mu\nu}(q) \,.
\ee
Clearly from Eq.(\ref{sd1}), 
we conclude that $\widehat\Delta^{-1}(0) = {\displaystyle\lim_{q^2\to 0}} (- q^2 B(q^2)) $.

Interestingly enough, 
once the transversality of $\widehat{\Pi}_{\mu\nu}$ has been
enforced,  the value of $\widehat{\Delta}^{-1}(0)$ is 
determined {\it solely} by ${\displaystyle\lim_{q^2\to 0}}\left(- q^2 B(q^2)\right)$.
Evidently, if $B(q^2)$ does not contain $(1/q^2)$ terms, one has that 
${\displaystyle\lim_{q^2\to 0}} (- q^2 B(q^2))= 0$, and therefore
$\widehat{\Delta}^{-1}(0)=0$, despite the fact that  
${\cal T}_{0}$ has been assumed to be non-vanishing. Thus,  
if the full three-gluon vertex $\g$ 
satisfies the WI of (\ref{3gl}), but does not contain 
poles, then the seagull contribution ${\cal T}_{0}\neq 0$ of graph ($a_2$)
will cancel exactly against 
analogous contributions contained in graph ($a_1$), 
forcing $\widehat{\Delta}^{-1}(0)= 0$.

We next proceed to study the SD of Eq.(\ref{sd1}). 
We will follow the methodology developed in \cite{Cornwall:1982zr} and linearize the
equation by resorting to the Lehmann representation,
together with a gauge-technique inspired Ansatz for the vertex ${\g}_{\nu\alpha\beta}$.
This approximation yields a more tractable form for the resulting SD equation,
which for the purposes of this preliminary analysis should suffice; 
of course, a non-linear study must eventually be carried out, and lies within our immediate plans.

To simplify the form of the vertex required,
we drop the longitudinal
parts of  $\widehat{\Delta}_{\mu\nu}$ {\it inside} the integrals,
using $\widehat{\Delta}_{\mu\nu}(k) = -i g_{\mu\nu} \widehat\Delta(k)$. 
Omitting these terms does not interfere with the transversality of 
the external $\widehat{\Delta}_{\mu\nu}(q)$ \cite{Aguilar:2006gr}. Then we obtain
\bea
\widehat{\Pi}_{\mu\nu}(q) &=& 
\frac{1}{2}\, C_A \, g^2 \,
 \int\!  [dk]\,
\widetilde{\Gamma}_{\mu}^{\alpha\beta}
\widehat{\Delta}(k) 
{\g}_{\nu\alpha\beta} 
\widehat{\Delta}(k+q)  \nonumber \\
 &-&\, C_A \, g^2 \, \,d \, g_{\mu\nu}
\int\!  [dk] \, \widehat{\Delta}(k) \, , 
\label{polar2}
\eea
with
\be
\widetilde\Gamma_{\mu\alpha\beta}= (2k+q)_{\mu}g_{\alpha\beta} -2q_\alpha g_{\mu\beta} + 
2 q_\beta g_{\mu\alpha} \, ,
\label{PTV}
\ee
and
\be
q^{\nu} {\g}_{\nu\alpha\beta} = 
\left[\widehat{\Delta}^{-1}(k+q) - \widehat{\Delta}^{-1}(k)\right] g_{\alpha\beta}\, .
\label{WID}
\ee 

The Lehmann representation for the scalar part of the gluon propagator reads
\begin{equation}
\widehat\Delta (q^2) = \int \!\! d \lambda^2 \, \frac{\rho\, (\lambda^2)}{q^2 - \lambda^2 + i\epsilon}\, ,
\label{lehmann}
\end{equation}
with no special assumptions on the form of the spectral density.

This way of writing $\widehat\Delta (q^2)$ allows for a relatively simple gauge-technique Ansatz for  ${\gt}$, which linearizes the resulting SDE.
In particular, setting on the first integral of the RHS of Eq.(\ref{polar2})
\begin{widetext}             %%%%%two columns
\begin{equation}
\widehat\Delta (k)\, {\g}_{\nu\alpha\beta}\,
\widehat\Delta (k+q) 
= \int \!\! d \lambda^2 \, \rho\,(\lambda^2)  
 \frac{1}{k^2 - \lambda^2 + i\epsilon} \,
{\gt} \,
\frac{1}{(k+q)^2 -  \lambda^2 + i\epsilon} \,,
\label{propvert}
\end{equation}
where ${\gt}$ must be such as to satisfy the tree-level WI
\be
q^{\nu} {\gt} = \left[(k+q)^2 - k^2 \right] g_{\alpha\beta} =\left[(k+q)^2 - \lambda^2 \right] g_{\alpha\beta} - (k^2 -\lambda^2)g_{\alpha\beta}\,. 
\label{wigtt}
\ee
Then, it is straightforward to show by contracting both sides 
of (\ref{propvert}) with $q^{\nu}$, and employing  
(\ref{lehmann}) and (\ref{wigtt}),  that ${\gt}$ satisfies the all-order WI of Eq.(\ref{WID}).
Of course, 
choosing ${\gt} = \widetilde\Gamma_{\nu\alpha\beta}$ solves the WI, but 
as we will see in detail in what follows, 
due to the absence of pole terms, it does
not allow for mass generation, in accordance with our previous discussion.
Instead we propose the following form for the vertex
\be
{\gt} =
\widetilde\Gamma_{\nu\alpha\beta}    +
    c_1 \left((2k+q)_{\nu} + \frac{q_{\nu}}{q^2}
\left[k^2 - (k+q)^2\right]\right)g_{\alpha\beta} 
%\nonumber\\
% && \hspace{-1cm}
 + \left( c_3 + \frac{c_2}{2\, q^2}\left[(k+q)^2 + k^2 \right]\right)
\left( q_\beta g_{\nu\alpha} - q_\alpha g_{\nu\beta} \right), \,
\label{vertpoles}
\ee 
which, due to the  presence of the massless pole is expected to 
allow the  possibility of infrared finite solution.  
Furthermore, we
treat  the  constants   $c_1$,  $c_2$  and  $c_3$  as  arbitrary
parameters, in order to  check quantitatively the sensitivity
of  the  obtained solutions  on  the   specific  details  of  the  form  of  the
vertex. Notice that all new terms contributing to ${\gt}$ 
have the correct properties under Bose symmetry.

  Thus,  the  vertex  ${\g}$  entering  in  Eq.(\ref{polar}),  can  be
  obtained    as   a    combination   of    Eqs.(\ref{propvert})   and
  (\ref{vertpoles}). After rather lengthy  algebraic manipulations of Eq.(\ref{sd1}) 
(see \cite{Aguilar:2006gr} for details), fixing $c_3=\frac{1}{3} c_1$ to recover the perturbative result, 
we obtain for the renormalized $\widehat\Delta (q^2)$
(in the Euclidean space) 
%
%\begin{widetext}
\be
{\widehat\Delta}^{-1}(q^2) = q^2 \Bigg\{ K
+ {\tilde b} \,g^2 
\int^{q^2/4}_{0}\!\!\!dz \, \left(1-\frac{4z}{q^2}\right)^{1/2}\,
{\widehat\Delta}(z)\Bigg\}
+  \, \gamma {\tilde b} g^2 
\int^{q^2/4}_{0}\!\!\!dz \,z \,\left(1-\frac{4z}{q^2}\right)^{1/2}
\widehat\Delta(z) 
\,\,+ {\widehat\Delta}^{-1}(0) \,, 
\label{sd4}
\ee
\end{widetext}                                 %%%%%two columns
with ${\tilde b}=10 \,C_A/48\pi^2 $,
\be
\widehat\Delta^{-1}(0) = - \frac{{\tilde b} g^2 \sigma}{\pi^2}
\int\! d^4 k \,\widehat\Delta (k^2) \, ,
\label{D02}
\ee 
and
\be
\sigma\,\equiv \, \frac{6\,(c_1+c_2)} {5}\,, \,\,\,\, \,\,\,\, \gamma\,\equiv \,\frac{4+4\,c_1+3\,c_2} {5}\,.
\label{coef}
\ee 
The renormalization constant $K$ is to be fixed 
by the condition, ${\widehat\Delta}^{-1}(\mu^2)=\mu^2$, with $\mu^2 \gg  \Lambda^2$.
Notice that the deviation of ${\tilde b}$ from the value $b=11 \,C_A/48\pi^2$, the 
standard coefficient of the one-loop $\beta$ function of QCD, is due to the omission of the ghosts.
From (\ref{D02}) it is clear that when $\sigma=0$ automatically $\widehat\Delta^{-1}(0)$ vanishes,  despite the inclusion of ($a_2$). Note however that having poles is not a sufficient condition: if $c_1=-c_2$, there is no effect. 

It is interesting to study the UV behavior for 
${\widehat\Delta}(q^2)$ predicted by the integral equation (\ref{sd4}).  
At large $q^2$ we can safely replace the factors  $(1- 4z/q^2)^{1/2}\to 1$,  
arriving at the following simplified version of that equation,
\begin{equation}
\widehat\Delta^{-1}(q^2) = q^2\left(1+ {\tilde b}g^2\int^{q^2}_{\mu^2}\!\!\! dz\,\widehat\Delta(z)  
\right)\,,
\label{uv}
\end{equation}
whose solution can be easily obtained by casting it into an 
differential equation, 
written in terms of the form factor $G(q^2)= q^2\widehat\Delta(q^2)$, which lead us to
\begin{equation}
\widehat\Delta^{-1}(q^2)= 
q^2\left[1+ 2 {\tilde b}g^2\ln\left(\frac{q^2}{\mu^2}\right)\right]^{1/2} \,.
\label{uv_sol}
\end{equation}
Obviously, upon expansion this expression recovers the one-loop result  
${\widehat\Delta}^{-1}(q^2)|_{pert}= q^2 \left(1+ {\tilde b} \,g^2 \ln(q^2/\mu^2)\right)$ correctly, 
but ${\widehat\Delta}(q^2)$ does not display the expected RG behavior at higher
order.  The fundamental 
reason for this discrepancy can be essentially traced back to having 
carried out the renormalization subtractively instead of multiplicatively 
\cite{Cornwall:1982zr,Cornwall:1985bg}, a fact that distorts the RG structure of the equation.

As is well-known,
due to the Abelian WI satisfied by the PT effective Green's functions, 
$\widehat\Delta^{-1}(q^2)$ absorbs all  
the RG-logs, exactly as happens in QED with the photon self-energy.
Consequently, the product 
${\widehat d}(q^2) = g^2 \widehat\Delta(q^2)$ should form a RG-invariant 
($\mu$-independent) quantity.
Notice however that Eq.(\ref{sd4}) does not encode the correct RG behavior: 
when written in terms of the RG invariant quantity 
$\widehat d(q^2) = g^2\widehat \Delta (q^2)$ it is not
manifestly $g^2$-independent, as it should.

In  order  to  restore  the  correct  RG  behavior  at  the  level  of
(\ref{sd4}),  observe that such  equation requires  an extra  power of
$g^2$  in  their  integrands on  the  RHS.  Then,  we use  the  simple
prescription whereby we  substitute every $\widehat\Delta(z)$ appearing
on RHS of Eq.(\ref{sd4}) by~\cite{Cornwall:1982zr,Cornwall:1985bg}
\be
\widehat\Delta(z) \to \frac{g^2\, \widehat\Delta(z)}{\bar{g}^{2}(z)}\equiv [1+ \tilde{b} g^2\ln(z/\mu^2)]\widehat\Delta(z) \,. 
\label{gratio1}
\ee
which allows us to cast Eq.(\ref{sd4}) in terms 
of the RG-invariant quantities  $\widehat d(q^2)$ and $\bar{g}^{2}(q^2)$ in the following way:
\begin{widetext}
\bea
{\widehat d}^{\,-1}(q^2) &=& q^2 \Bigg\{ \frac{1}{g^2}
+ {\tilde b}\Bigg( 
\int^{q^2/4}_{0}\! dz\, \left(1-\frac{4z}{q^2}\right)^{1/2}
\frac{{\widehat d}(z)}{\bar{g}^2(z)}
- 
\int^{\mu^2/4}_{0}\!dz \, 
\left(1-\frac{4z}{\mu^2}\right)^{1/2}\,\frac{{\widehat d}(z)}{\bar{g}^2(z)} \Bigg)\Bigg\}
\nonumber\\
&+&\,
\gamma {\tilde b}\Bigg(
\int^{q^2/4}_{0}\!dz \,z \,
\left(1-\frac{4z}{q^2}\right)^{1/2}
\frac{{\widehat d}(z)}{\bar{g}^2(z)} \,\,
- \frac{q^2}{\mu^2} \int^{\mu^2/4}_{0}\!dz  \, z  
\left(1-\frac{4z}{\mu^2}\right)^{1/2}\,\frac{{\widehat d}(z)}{\bar{g}^2(z)}\Bigg)
+
{\widehat d}^{\,-1}(0) \;,
\label{sd6}
\eea
\end{widetext}
where 
\be
{\widehat d}^{\,-1}(0) = 
 -  \frac{{\tilde b}\sigma}{\pi^2}
\int\! d^4 k \,\frac{{\widehat d} (k^2)}{\bar{g}^2(k^2)} \,.
\label{d01}
\ee 
It is easy to see now that Eq.(\ref{sd6}) yields the correct UV behavior, i.e.  ${\widehat d}^{\,-1}(q^2)= \tilde{b}\,q^2\ln(q^2/\Lambda^2)$.

When solving (\ref{sd6})
we will be interested in solutions that are qualitatively of the general form  
\be
{\widehat d}(q^2) = \frac{\gnp(q^2)}{q^2 + m^2(q^2)}\,,
\label{ddef}
\ee
where 
\be
\gnp(q^2) = \bigg[\tilde{b}\ln\left(\frac{q^2 + f(q^2, m^2(q^2))}{\Lambda^2}\right)\bigg]^{-1}\,,
\label{GNP}
\ee
The quantity $\gnp(q^2)$
represents a non-perturbative version of the 
RG-invariant effective charge of QCD:
in the deep UV it goes over to $\overline{g}^{2}(z)$, 
while in the deep IR it will be finite 
due to the presence of the function 
$f(q^2, m^2(q^2))$, whose form will be determined by fitting the numerical solution.  

The function $m^2(q^2)$ may be interpreted as a  
momentum dependent ``mass''. On general arguments dynamically generated masses must 
vanish asymptotically. In order
to determine the  asymptotic behavior that Eq.(\ref{sd6}) predicts
for  $m^2(q^2)$ at large $q^2$,
we replace Eq.(\ref{ddef}) on both sides of Eq.(\ref{sd6}), set
$(1- 4z/q^2)^{1/2}\to 1$, and demand the consistency of both sides, obtaining finally 
\be
 m^2(q^2) \sim  m^2_{0} \ln^{-a} \left(q^2/\Lambda^2\right)\,, \quad\mbox{with}\quad a= 1+\gamma  \, . 
\label{uv_mass} 
\ee 
Indeed the  gluon mass vanishes at UV as an inverse
power  of $\ln(q^2)$,  since $a>0$.   Actually,  as we  will see
below,  the regularization  of  Eq.(\ref{d01}) imposes a more stringent 
constraint, requiring that $\gamma > 0$, thus restricting
through Eq.(\ref{coef}) the possible values of $c_1$ and $c_2$ in ${\gt}$.

As  mentioned before, the  seagull-like contribution  
(denoted collectively by ${\widehat d}^{\,-1}(0)$ in Eq.(\ref{sd6})) 
are essential for obtaining IR   finite    solution   for   ${\widehat
d}(q^2)$.  However,  the   integral  (\ref{d01})  should  be  properly
regularized, in order to ensure the finiteness of such a mass term.

  For the regulation of the quadratic  divergences present  in the integral
  (\ref{d01}),  we rely  on two  basic ingredients:  (i)  the standard
  integration rules of the  dimensional regularization and (ii) a constraint on
  the allowed values of the anomalous mass-dimension $a$.
  
 With this in mind, we recall that according to the dimensional regularization rules, $\int\!\,[dk]/k^2 =0$, allowing us to rewrite the Eq.(\ref{d01}) (using (\ref{ddef})) as  
%\begin{widetext}
\bea
{\widehat d}^{\,-1}(0) &\equiv&   -  \frac{{\tilde b}\sigma}{\pi^2}
\int\! [dk] \bigg(\,\frac{\gnp(k^2)}{[k^2 + m^2(k^2)]\bar{g}^{2}(k^2)} -\frac{1}{k^2}\bigg)
\nonumber\\
&& \hspace{-1cm}=  \frac{{\tilde b}\sigma}{\pi^2} \int\! [dk] \frac{m^2(k^2)}{k^2\, [k^2 + m^2(k^2)]}\, \nonumber\\ 
&& \hspace{-1cm}+  \frac{{\tilde b^2}\sigma}{\pi^2} \int\! [dk]\, {\widehat d}(k^2)\,
\ln\left(1 + \frac{f(k^2, m^2(k^2))}{k^2}\right) \, .
\label{basreg}
\eea
%\end{widetext}

The inspection of the two integrals on the RHS separately reveals that, 
if $m^2(k^2)$ falls asymptotically as power of $\ln^{-a} (k^2)$,  
with $a >1$, then the first integral would  converge,
by virtue of the elementary result 
\be
\int  \frac{dz}{z\, (\ln z)^{1+\gamma}} = - \frac{1}{\gamma \, (\ln z)^{\gamma}}\,,
\label{elint}
\ee
which, of course, requires that $\gamma >0$.
The second integral will converge as well, provided that $f(k^2, m^2(k^2))$ drops asymptotically at least as fast as $\ln^{-c} (k^2)$, with $c>0$. If, for example, $f=\rho m^2(k^2)$ (with $1+\gamma >1$,
for the first integral to converge),
then the convergence condition for the second integral is automatically fulfilled.  Notice that perturbatively $ {\widehat d}^{\,-1}(0)$ 
vanishes; this is because   $m^2(k^2)=0$ to all orders, 
and therefore, since in that case also $f=0$,
both integrals on the RHS of (\ref{basreg}) vanish.

Evidently, Eqs.(\ref{sd6}) and (\ref{basreg}) form  a system of equation;
the role of   the first is to provide a solution  for the unknown RGI quantity, 
${\widehat d}(q^2)$, while the second acts as an additional constraint, restricting the number of 
possible solutions.
Therefore,  Eqs.(\ref{sd6}) and (\ref{basreg}) should be solved simultaneously. 

Using an iterative method, we performed a detailed study of these two equations, 
where for each  ${\widehat d}^{-1}(0)$ chosen 
we  vary $\gamma$ and $\sigma$ in order to scan the two-parameter space of solutions.

\begin{figure}[ht]
\vspace{-1cm}
\hspace{-1.0cm}
\includegraphics[scale=0.9]{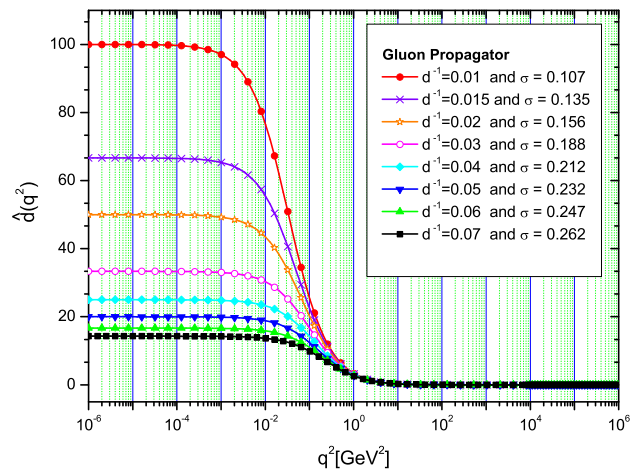}
\vspace{-1cm}
\caption{Results for ${\widehat d}(q^2)$ fixing different values for ${\widehat d}^{\,-1}(0)$ (all in  $\;\mbox{GeV}^{\,2}$). All these solutions satisfy the condition given by Eq.(\ref{basreg}). Their respective values for $\sigma$ are given in the legend, in all cases we set $c_2=0$ in Eq.(\ref{coef}).}
\label{p1}
\end{figure}

In Fig.(\ref{p1}), we show  numerical results for ${\widehat d}(q^2)$,
for different values of  ${\widehat d}^{\,-1}(0)$. All these solutions
satisfy  the   constraint  imposed  by   Eq.(\ref{basreg})  and  their
respective  values for  $\sigma$   are  described in  the
inserted  legend. As  expected, the  gluon propagator  behaves  at low
momenta  like a  constant,  whose value  is  determined by  ${\widehat
d}(0)$; in addition, it obeys the correct ultraviolet behavior. As can
be observed from the plot,  ${\widehat d}(q^2)$ starts off as constant
until  the neighborhood  of $q^2=0.01  \;\mbox{GeV}^{\,2}$,  where the
curve  bends  downward  in   order  to  match  with  the  perturbative
asymptotic behavior at a scale of a few $\mbox{GeV}^{\,2}$.  All these
solutions can  be perfectly fitted by  Eq.(\ref{ddef}); the functional
form  of $f(q^2, m^2(q^2))$  and $m^2(q^2)$  that better  describe our
data sets are given by
\be
f(q^2, m^2(q^2)) = \rho_{\,1} m^2(q^2)+ \rho_{\,2} \frac{m^4(q^2)}{q^2+m^2(q^2)} \,,
\label{func_fit}
\ee
and the dynamical mass, 
\be
m^2(q^2)=m^2_0\left[\ln
\left[\frac{q^2+\rho_{\,1}\,m^2_0}{\Lambda^2}\right]\Big/\ln\left[\frac{\rho_{\,1}\,m^2_0}{\Lambda^2}\right] \right]^{-a} \,,
\label{dmass}
\ee 
with exponent $a = 1+ \gamma $.%   

In  all cases,  we fixed  $\rho_1=4$; therefore, the  unique free
parameter is $\rho_{\,2}$,  since the value of $m^2_0$,  which is also
related to $\rho_{\,2}$, can  be directly obtained by setting $q^2=0$
in Eq.(\ref{ddef}). From this follows immediately that the value of the
infrared   fixed   point   of   the  running   coupling,   ${\overline
g}^{\,-2}_{{\chic  {\rm  NP}}}(0)$ will  be  determined  by the  value
assumed by $f(0,m^2_0)$, since
\be
{\overline g}^{\,-2}_{{\chic {\rm NP}}}(0) = \tilde{b}\ln\left[\frac{(4 +\rho_{\,2})\,m^2_0 }{\Lambda^2}\right] \,.
\label{fita1}
\ee
% 
%%%%%%%%%%%%%%%%%%%%%%%% FIGURE 2 %%%%%%%%%%%%%%%%%%%%%%%%%%%%%%%%%%%%%
%    
%%%%%%%%%%%%%%%%%%%%%%%%%%%%%%%%%%%%%%%%%%%%%%%%%%%%%%%%%%%%%%%%%%%%
%%%%%%%%%%%%%%%%%%%%%%%%%%%%%%%%%%%%%%%%%%%%%%%%%%%%%%%%%%%%%%%%%%%%%%%
%                               fig3
%%%%%%%%%%%%%%%%%%%%%%%%%%%%%%%%%%%%%%%%%%%%%%%%%%%%%%%%%%%%%%%%%%%%
\begin{figure}[ht]
\vspace{-1cm}
\hspace{-1.3cm}
\includegraphics[scale=0.9]{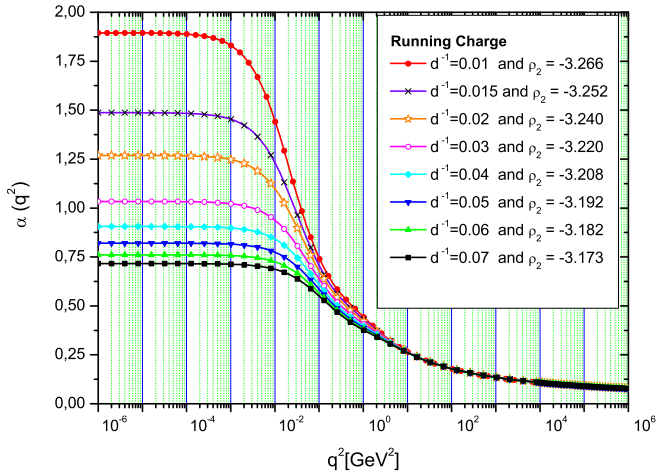}
\vspace{-1cm}
\caption{The running charge, $\alpha(q^2)$, corresponding 
to the gluon propagator of Fig.(\ref{p1}).  Clearly, 
$\alpha(0)$ increases as ${\widehat d}^{\,-1}(0)$ decreases}
\label{p2}
\end{figure}
%%%%%%%%%%%%%%%%%%%%%%%%%%%%%%%%%%%%%%%%%%%%%%%%%%%%%%%%%%%%%%%%%%%%%%%
Obviously, the maximum value obtained for $\gnp(0)$, is the one that
minimizes $(4  +\rho_{\,2})\,m^2_0$ and, at  same time, keeps  it bigger
than  $\Lambda^2$, in  order to  avoid  the pole.  Of course,  if we  set
$\rho_{\,2}=0$  in Eq.(\ref{func_fit}),  we automatically  recover the
solution  proposed in  \cite{Cornwall:1982zr};  however our  numerical
solution  requires bigger  values  for the  coupling $\gnp(q^2)$,  and
therefore $\rho_{\,2}$ assumes negative values, as can be observed on
Fig.(\ref{p2})

The  running charges,  $\alpha(q^2)$, for  each solution  presented on
Fig.(\ref{p1})  are  displayed  in  Fig.(\ref{p2}). Observe  that as
the value of  ${\widehat d}^{\,-1}(0)$ decreases, the
value of the infrared    fixed    point    of    the    running    coupling,
$\alpha(0)$, increases. Accordingly,  from Fig.(\ref{p1}) and  Fig.(\ref{p2}), we
can conclude that, if  smaller values of $\sigma$ were to be favored
by QCD,  the freezing  of the running  coupling would occur  at higher
values. It should also be noted that the values of $\alpha(0)$ found
here tend to be slightly more elevated compared to those of 
\cite{Cornwall:1982zr} (for the same value of $m^2_0$) .

\begin{figure}[t!]
\vspace{-1cm}
\hspace{-0.9cm}
\includegraphics[scale=0.9]{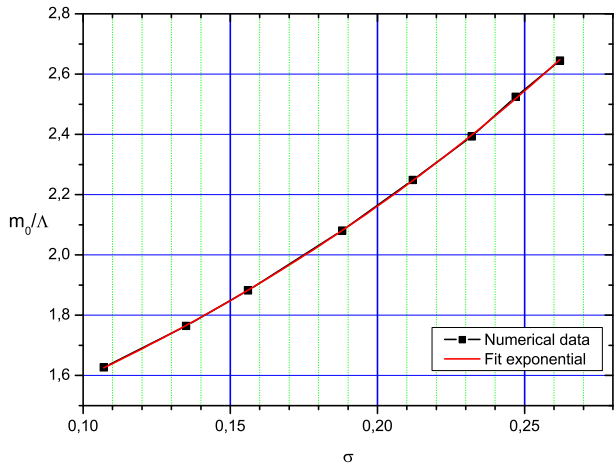}
\vspace{-1cm}
\caption{The ratio $m_0/\Lambda$ as function of the parameter $\sigma$. 
The increase is exponential, given by
$m_0/\Lambda = A_1\exp(\sigma/t_1) + y_0$, where $A_1=0.775$, $t_1=0.25$ and $y_0=0.436$.}
\label{p3}
\end{figure}

Finally, we analyze the dependence
of the ratio $m_0/\Lambda$ on $\sigma$; the former is  
extracted from Eq.(\ref{ddef}) by setting  $q^2=0$.
This dependence  is shown  in the
Fig.(\ref{p3}), corresponding to the cases  presented in Fig.(\ref{p1}).
We  observe that  as we  increase the  value of
$\sigma$,   namely the sum of the coefficients of 
the  massless  pole  terms  appearing  in  the three  gluon  vertex,
the  ratio   $m_0/\Lambda$  grows  exponentially.

%
%%%%%%%%%%%%%%%%%%%%%%%% plot 4 %%%%%%%%%%%%%%%%%%%%%%%%%%%%%%%%%%%%%
%\begin{figure}[ht]
%%%%%%%%%%%%%%%%%%%%%%%%%%%%%%%%%%%%%%%%%%%%%%%%%%%%%%%%%%%%%%%%%%% 

 In  conclusion,  we  have   presented  an  analysis  of  the  various
intertwined  issues involved  in  the study  of  dynamical gluon  mass
generation through SD equations.  The  analysis was carried out in the
context  of the PT-BFM  scheme, where  various crucial  properties are
preserved  manifestly.   Most  notably,  the   transversality  of  the
non-perturbative gluon  self-energy is enforced order by  order in the
dressed loop expansion  and separately for gluons and  ghost.  We have
seen in detail that for  the existence of infrared finite solutions two
requirements  are  {\it   indispensable}:  (i)  the  non-vanishing  of
seagull-like  contributions beyond perturbation  theory, and  (ii) the
presence  of  massless  poles  in  the trilinear  gluon  vertex.   The
resulting  equation  was  linearized   by  resorting  to  the  Lehmann
representation and  a   gauge-technique  inspired  solution  of  the
corresponding  WI.   A  simple   Ansatz  for  the three-gluon  vertex  was
constructed,  which contains  massless  poles, thus  allowing for  the
appearance of  infrared finite  solutions.  This vertex  satisfies the
correct  WI, but otherwise  is purely  phenomenological, in  the sense
that  it is  not  QCD-derived, nor  does  it contain  the most  general
Lorentz  structure.    Numerical  solutions  were   obtained  for  the
RG-invariant  quantity   ${\widehat  d}(q^2)$;  they   can  be  fitted
perfectly by means of a running coupling that freezes in the IR, and a
dynamical mass that vanishes in the UV.  We have found that the actual
values of $\alpha(0)$  depend strongly on the combined  strength of the
pole terms appearing  in the vertex.  This strongly  suggests that the
value of this  IR fixed point should be determined  by means of a
detailed non-perturbative  study of the three-gluon  vertex, either on
the  lattice or through  its own SD equation.   Needless to
say, many of the issues considered  in this talk are far from settled,
and a lot of independent  work is  necessary before  reaching definite
conclusions.

\vspace{-0.5cm}

\section*{Acknowledgments}
A.C.A. thanks the organizers  of IRQCD for their hospitality. Research
supported by FAPESP/Brazil through  the grant  (05/04066-0)  and by  the
Spanish MEC under the Grant FPA 2005-01678.

\end{document}